\documentclass[iop]{emulateapj}
\usepackage{natbib,graphicx,epstopdf}
\newcommand{\teff}{$T_\mathrm{eff}$}

\newcommand{\caii}{\ion{Ca}{2} H \& K}
\newcommand{\logr}{log(\ensuremath{R'_{\mbox{\scriptsize HK}}})}
\newcommand{\vsini}{$v \sin i$}
\newcommand{\sval}{\ensuremath{S_{\mbox{\scriptsize HK}}}}
\newcommand{\lya}{Lyman $\alpha$}

\shorttitle{A Correlation Between Stellar Activity and Hot Jupiter Emission Spectra}
\shortauthors{Knutson et al.}\def\simgr{\,\hbox{\hbox{$ > $}\kern -0.8em \lower 1.0ex\hbox{$\sim$}}\,}
\def\simle{\,\hbox{\hbox{$ < $}\kern -0.8em \lower 1.0ex\hbox{$\sim$}}\,}

\begin{document}

\title{A Correlation Between Stellar Activity and Hot Jupiter Emission Spectra\altaffilmark{1}}

\author{
Heather A. Knutson\altaffilmark{2,3},
Andrew W.\ Howard\altaffilmark{2,4}, 
\&~Howard Isaacson\altaffilmark{2}
}
\altaffiltext{1}{Based on observations obtained at the W.\,M.\,Keck Observatory, 
                      which is operated jointly by the University of California and the 
                      California Institute of Technology.  Keck time has been granted by both 
                      NASA and the University of California.} 
\altaffiltext{2}{Department of Astronomy, University of California, Berkeley, CA 94720-3411, USA} 
\altaffiltext{3}{Miller Fellow; hknutson@berkeley.edu}
\altaffiltext{4}{Townes Fellow, Space Sciences Laboratory, University of California, 
                        Berkeley, CA 94720-7450 USA}

\begin{abstract}

We present evidence for a correlation between the observed properties of hot Jupiter emission spectra and the activity levels of the host stars measured using \caii~emission lines.  We find that planets with dayside emission spectra that are well-described by standard 1D atmosphere models with water in absorption (HD 189733, TrES-1, TrES-3, WASP-4) orbit chromospherically active stars, while planets with emission spectra that are consistent with the presence of a strong high-altitude temperature inversion and water in emission orbit quieter stars.  We estimate that active G and K stars have \lya~fluxes that are typically a factor of $4-7$ times higher than quiet stars with analogous spectral types, and propose that the increased UV flux received by planets orbiting active stars destroys the compounds responsible for the formation of the observed temperature inversions.  In this paper we also derive a model-independent method for differentiating between these two atmosphere types using the secondary eclipse depths measured in the 3.6 and 4.5~\micron~bands on the \emph{Spitzer Space Telescope}, and argue that the observed correlation is independent of the inverted/non-inverted paradigm for classifying hot Jupiter atmospheres.  

\end{abstract}


\section{Introduction}

The close-in, gas giant planets known as ``hot Jupiters" occupy a unique regime of parameter space, with temperatures between $1000-2500$~K and atmospheric compositions that are likely to be quite different than that of the cooler solar system gas giants.  Because the time scale for tidal synchronization is short compared to the ages of these systems, hot Jupiters in circular orbits are expected to be tidally locked, leaving one side of the planet in permanent darkness while the other is constantly illuminated by intense radiation from the star.  This further complicates the atmospheric chemistry, as the cooler night side may act as either a source or a sink for different materials, while vigorous day-night circulation may produce increased vertical mixing on the dayside.  These planets also experience much higher UV fluxes than the solar system gas giants, and it is likely that photochemistry alters the atmospheric chemistry down to pressures of 10 mbars or more \citep{liang03,liang04,zahnle09,zahnle10,line10}.  Infrared light from these planets emerges from pressures of 10-100 mbar \citep[e.g.][]{burrows08,showman09}, and as a result their dayside emission spectra could exhibit strong features from photochemically produced molecules.  Indeed, the detection of an excess of CO$_2$ in the atmosphere of the hot Jupiter HD 189733b \citep{swain09a} indicates that non-equilibrium chemistry may play an important role in these atmospheres.

\begin{figure}
\epsscale{1.0}
\plotone{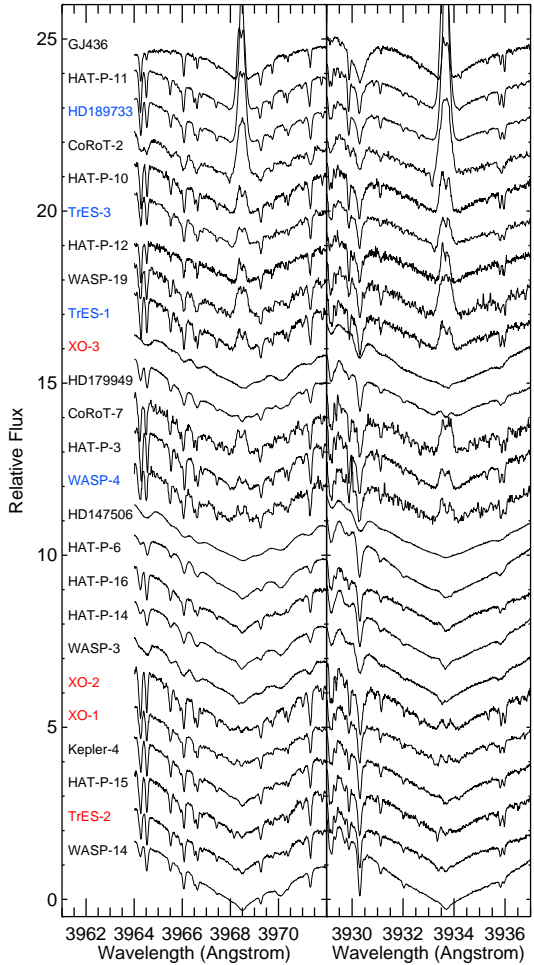}
\caption{\caii~lines for transiting planet host stars observed with Keck/HIRES, plotted in order of increasing \sval.  This plot shows half of our sample, with the second half plotted in Fig. \ref{spectrab}.  Active stars have significant emission in the line cores (high values of \sval) and lie at the top of the plot; a majority of the stars in our sample appear to have no detectible emission.  Star names are given on the left, with systems hosting planets with temperature inversions shown in red and those without temperature inversions shown in blue, while names in black denote systems with insufficient data.  It is possible that cooler ($<1000$~K) and/or higher density planets, such as the Neptune-mass planets GJ 436 and HAT-P-11, may not fit into the simple classification scheme used to describe hot Jupiters atmospheres \citep[indeed this appears to be the case for GJ 436;][]{stevenson10}, but we include HIRES observations in this plot where available for completeness.}
\label{spectraa}
\end{figure}

\begin{figure}
\epsscale{1.0}
\plotone{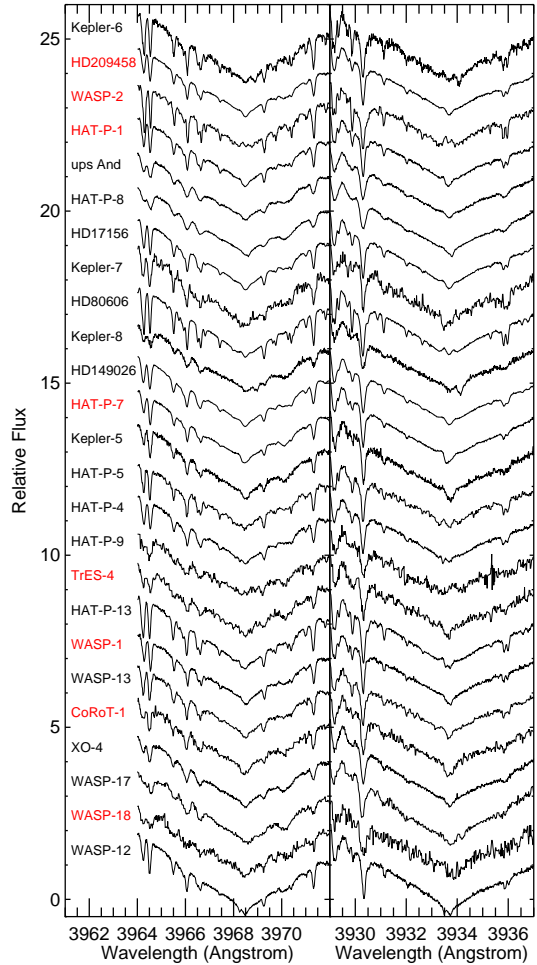}
\caption{\caii~lines for transiting planet host stars observed with Keck/HIRES, plotted in order of increasing \sval.  This plot is a continuation of Fig. \ref{spectraa}, and includes stars with lower values of \sval.}
\label{spectrab}
\end{figure}

We can characterize the dayside emission spectra of these planets by measuring the wavelength-dependent decrease in light as the planet passes behind the star in an event known as a secondary eclipse.  Secondary eclipse observations have now been carried out for more than sixteen extrasolar planets, and there is currently a growing body of evidence for the existence of two distinct classes of hot Jupiter atmospheres  \citep[see][and references therein]{deming09}.  According to the current classification scheme one class of planets \citep[e.g. HD 189733;][]{charbonneau08} have dayside emission spectra that are well-described by atmosphere models with water and CO in absorption.  A second class of planets \citep[e.g. HD 209458;][]{knutson08} are better described by models with a strong temperature inversion between 0.1-0.01 bars and these same features in emission.  

We do not know the nature of the absorber required to create these inversions; it was initially suggested that gas-phase TiO might provide the necessary opacity\citep{hubeney03,burrows07,burrows08,fortney06,fortney08}, but this theory cannot explain the full range of current observations.  TrES-3 is hot enough for gas-phase TIO but does not appear to have an inversion \citep{fressin10}, whereas XO-1 receives a much lower incident flux comparable to that of HD 189733b and yet has an inversion \citep{machalek08}.  It is also unclear whether or not TiO can be reliably maintained in the upper atmosphere at all, given its high molecular weight and the likely presence of day- and night-side cold traps.  Overcoming these effects would require vigorous mixing, but this may not be expected in a stably stratified atmosphere \citep{spiegel09}.  More recently, \citet{zahnle09} presented detailed non-equilibrium atmospheric chemistry models suggesting that heating from sulfur compounds in the upper atmospheres of hot Jupiters could explain these inversions; in this model, photochemistry serves as a net sink for the sulfur compounds of interest.  

In order to quantify the effects of photochemistry in these atmospheres more precisely, we must first obtain an estimate of the incident UV flux.  Direct measurements of the stellar UV fluxes are only available for a handful of systems, including HD 189733b \citep{lecavelier10} and HD 209458b \citep{vidal03,vidal04,ehrenreich08,france10,linsky10}, while archival searches for X-ray detections \citep{kashyap08} and more recent observations of planet-hosting stars with XMM-Newton \citep{poppen10} provide only upper limits in most cases.  In this paper we present measurements of \caii~line strengths \citep{noyes84}, which act as an indicator of stellar activity levels, based on Keck HIRES spectra for fifty transiting planet host stars.  These observations allow us to identify the most active planet-hosting stars, which should have correspondingly enhanced UV and X-ray fluxes \citep[e.g.][]{schrijver92,livingston07}, and to search for evidence that increased activity in the host star can explain the observed dichotomy in hot Jupiter emission spectra.  We describe our observations in \S\ref{obs} and survey the available literature on classifications of hot Jupiter emission spectra in \S\ref{sec_eclipses}.  In \S\ref{empirical_rel} we develop a model-independent metric for distinguishing between the two observed types of planetary atmospheres, and argue that planets without inversions are consistently found around the most active stars, while planets with strong inversions orbit quiet stars.  We then estimate the effect that increased activity has on the measured \lya~fluxes from the stars in our sample in \S\ref{uv}, and discuss the consequences this has for our understanding of hot Jupiter atmospheres in \S\ref{discussion}.  

\section{Keck/HIRES \caii~Observations}\label{obs}

We observed the transiting planet host stars listed in Table \ref{ca_hk_table} with the HIRES echelle spectrometer \citep{vogt94} on the 10-m Keck I telescope (see Fig. \ref{spectraa} and \ref{spectrab}).  We used the standard HIRES setup employed by the California Planet Search (CPS) group \citep{wright04,howard09}.  Most observations were made with a slit of width 0\mbox{\ensuremath{.\!\!^{\prime\prime}}}86 for 500\,s yielding a signal to noise of 75--200 per pixel (depending on the host star brightness).  Many of the observations were initially made for radial velocity monitoring, and additional single-exposure observations were obtained where necessary to expand this coverage to a larger sample of transiting planet host stars.  For the radial velocity observations, an iodine cell was mounted directly in front of the spectrometer entrance slit to provide a wavelength scale and calibration of the instrumental profile \citep{marcy92,valenti95}.  The iodine absorption lines (5000-6200\,\AA) do not affect the measurement of \caii~indices (3968\,\AA~and 3933\,\AA).  Stellar parameters were determined using a LTE spectral synthesis routine  \citep[SME or Spectroscopy Made Easy;][]{fischer05} on observations taken without the iodine cell where available.  Systems with only a single spectrum were observed in this mode.

\begin{figure}
\epsscale{1.1}
\plotone{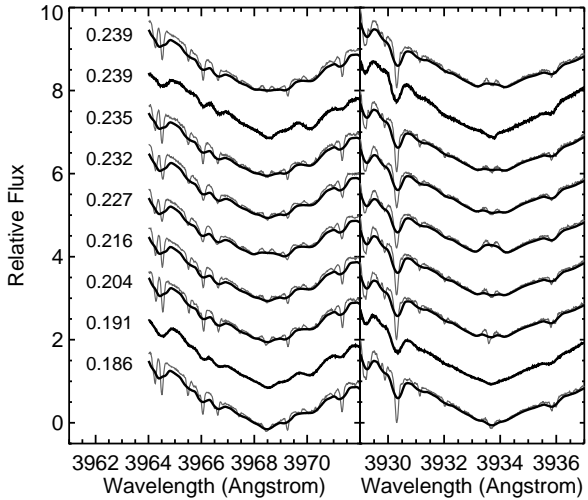}
\caption{\caii~lines for a sample of seven F stars observed with Keck/HIRES as part of ongoing CPS programs, plotted in order of increasing \sval~(values given on left).  These stars were selected to have effective temperatures between $6000-6500$~K, \vsini~less than 10 km s$^{-1}$, and \sval~values in the range $0.18-0.24$, comparable to the values measured for XO-3 (\sval=0.239) and HD 147506 (\sval=0.191).  The original spectra (thin grey lines) for these stars were then artificially broadened to a \vsini~ of 20 km s$^{-1}$ (thick black lines), comparable to that of XO-3 (second from top) and HD 147506 (second from bottom) which are plotted for comparison.  The effective temperatures for the stars shown in this plot are, from bottom to top, 6170, 6080, 6260, 6090, 6130, 6090, 6190, 6430, and 6090~K, respectively.}
\label{vsini_spectra}
\end{figure}

We calibrate the stellar activity indices to the Mt. Wilson $S$-value scale, defined as the ratio of the sum of the flux in the cores of the \caii~lines to the sum of two continuum bands, one redward and one blueward of the H and K lines \citep{wilson68}.  The calibration from flux to \sval~was accomplished using stars that were observed in both the Mt. Wilson H-K project \citep{duncan91} and the California Planet Search.  \sval~values for a particular star on the CPS scale are typically within 10\% of the Mt. Wilson scale \citep{isaacson10}.  Before extracting the H line at 3968.5\,\AA, the K line at 3933.7\,\AA~and continuum flux levels, the spectra  are shifted to rest wavelengths through a cross correlation of the NSO solar atlas. The flux in the H and K lines cores is measured in two 1\,\AA~FWHM weighted triangles centered on the H and K lines. The continuum sections are each 20\,\AA~wide and reside sufficiently far away from the H and K line centers so as to avoid the broad wings of either line.  After calculating the S-value for each star, we convert to \logr~using the $\bv$ color of the star  \citep{noyes84}; this effectively removes any dependencies on the bolometric flux of the star and allows us to accurately compare the chromospheric emission from different spectral types.  In cases where we have multiple spectra available spanning several epochs, we take the median value of \logr~over all of the available observations.  We list the resulting \sval~and \logr~values in Table \ref{ca_hk_table}; for a more detailed description of our observations and methodology see \citet{isaacson10}.

\begin{deluxetable}{lrrrrcrrrrr}
\tablecaption{\caii~Line Strengths \label{ca_hk_table}}
\tablewidth{0pt}
\tablenum{1}
\label{stellar_params}
\tablehead{
\colhead{Star} & \colhead{\teff \tablenotemark{b} (K)}  & \colhead{\vsini \tablenotemark{b}} & \colhead{\bv} & \colhead{\sval\tablenotemark{f}} & \colhead{\logr \tablenotemark{f}}} 
\startdata
HD 17156 & 5980\phantom{c} & 4.3 & 0.64\phantom{d} & 0.155\phantom{e} & -5.022\\
HD 147506 & 6260\phantom{c} & 20.7 & 0.41\phantom{d} & 0.191\tablenotemark{e} & -4.780\\
HD 149026 & 6150\tablenotemark{c} & 6.2 & 0.61\phantom{d} & 0.152\phantom{e} & -5.030\\
HD 179949 & 6170\phantom{c} & 7.0 & 0.50\phantom{d} & 0.232\phantom{e} & -4.622\\
HD 189733 & 5090\phantom{c} & 2.7 & 0.93\phantom{d} & 0.508\phantom{e} & -4.501\\
HD 209458 & 6070\phantom{c} & 4.0 & 0.59\phantom{d} & 0.160\phantom{e} & -4.970\\
HD 80606 & 5510\phantom{c} & 2.4 & 0.77\phantom{d} & 0.154\phantom{e} & -5.061\\
GJ 436 & 3590\tablenotemark{c} & $<3$ & 1.52\phantom{d} & 0.620\phantom{e} & -5.298\\
$\upsilon$~And & 6160\tablenotemark{c} & 9.7 & 0.54\phantom{d} & 0.156\phantom{e} & -4.982\\
TrES-1 & 5300\phantom{c} & 0.5 & 0.78\phantom{d} & 0.244\phantom{e} & -4.738\\
TrES-2 & 5840\phantom{c} & 1.6 & 0.62\phantom{d} & 0.165\phantom{e} & -4.949\\
TrES-3 & 5530\phantom{c} & 3.1 & 0.71\phantom{d} & 0.305\phantom{e} & -4.549\\
TrES-4 & 6200\tablenotemark{c} & 8.5 & 0.52\phantom{d} & 0.141\phantom{e} & -5.104\\
XO-1 & 5750\tablenotemark{c} & 1.1 & 0.69\phantom{d} & 0.168\phantom{e} & -4.958\\
XO-2 & 5340\tablenotemark{c} & 1.4 & 0.82\phantom{d} & 0.173\phantom{e} & -4.988\\
XO-3 & 6430\tablenotemark{c} & 18.5 & 0.46\tablenotemark{d} & 0.239\tablenotemark{e} & -4.595\\
XO-4 & 6400\tablenotemark{c} & 8.8 & 0.47\tablenotemark{d} & 0.124\phantom{e} & -5.292\\
HAT-P-1 & 5980\phantom{c} & 2.2 & 0.58\tablenotemark{d} & 0.158\phantom{e} & -4.984\\
HAT-P-3 & 5170\phantom{c} & 0.5 & 0.87\tablenotemark{d} & 0.206\phantom{e} & -4.904\\
HAT-P-4 & 5990\phantom{c} & 5.1 & 0.58\tablenotemark{d} & 0.145\phantom{e} & -5.082\\
HAT-P-5 & 5960\tablenotemark{c} & 2.6 & 0.59\tablenotemark{d} & 0.148\phantom{e} & -5.061\\
HAT-P-6 & 6410\phantom{c} & 8.5 & 0.41\phantom{d} & 0.187\phantom{e} & -4.799\\
HAT-P-7 & 6350\phantom{c} & 3.8 & 0.44\phantom{d} & 0.150\phantom{e} & -5.018\\
HAT-P-8 & 6130\phantom{c} & 11.5 & 0.54\tablenotemark{d} & 0.156\phantom{e} & -4.985\\
HAT-P-9 & 6350\tablenotemark{c} & 11.9 & 0.48\tablenotemark{d} & 0.141\phantom{e} & -5.092\\
HAT-P-10\tablenotemark{a} & 4990\phantom{c} & 0.5 & 1.01\phantom{d} & 0.322\phantom{e} & -4.823\\
HAT-P-11 & 4820\phantom{c} & 0.5 & 1.02\phantom{d} & 0.580\phantom{e} & -4.567\\
HAT-P-12 & 4650\phantom{c} & 0.5 & 1.13\tablenotemark{d} & 0.253\phantom{e} & -5.104\\
HAT-P-13 & 5710\phantom{c} & 2.7 & 0.73\phantom{d} & 0.141\phantom{e} & -5.138\\
HAT-P-14 & 6530\phantom{c} & 8.5 & 0.42\phantom{d} & 0.175\phantom{e} & -4.855\\
HAT-P-15 & 5570\tablenotemark{c} & 2.0 & 0.71\tablenotemark{d} & 0.166\phantom{e} & -4.977\\
HAT-P-16 & 6160\tablenotemark{c} & 3.5 & 0.53\tablenotemark{d} & 0.175\phantom{e} & -4.863\\
WASP-1 & 6170\phantom{c} & 0.2 & 0.53\tablenotemark{d} & 0.140\phantom{e} & -5.114\\
WASP-2 & 5230\phantom{c} & 1.3 & 0.84\tablenotemark{d} & 0.159\phantom{e} & -5.054\\
WASP-3 & 6170\phantom{c} & 14.2 & 0.52\phantom{d} & 0.173\phantom{e} & -4.872\\
WASP-4 & 5500\tablenotemark{c} & 2.0 & 0.74\tablenotemark{d} & 0.194\phantom{e} & -4.865\\
WASP-12 & 6300\phantom{c} & 3.0 & 0.50\tablenotemark{d} & 0.113\phantom{e} & -5.500\\
WASP-13 & 5910\phantom{c} & 5.0 & 0.60\tablenotemark{d} & 0.127\phantom{e} & -5.263\\
WASP-14 & 6270\phantom{c} & 3.9 & 0.45\phantom{d} & 0.163\phantom{e} & -4.923\\
WASP-17 & 6380\phantom{c} & 10.7 & 0.48\tablenotemark{d} & 0.121\phantom{e} & -5.331\\
WASP-18 & 6250\phantom{c} & 11.7 & 0.49\phantom{d} & 0.116\phantom{e} & -5.430\\
WASP-19 & 5590\phantom{c} & 5.3 & 0.70\tablenotemark{d} & 0.252\phantom{e} & -4.660\\
CoRoT-1 & 5950\tablenotemark{c} & 5.2 & 0.59\tablenotemark{d} & 0.124\phantom{e} & -5.312\\
CoRoT-2 & 5630\tablenotemark{c} & 11.5 & 0.69\tablenotemark{d} & 0.435\phantom{e} & -4.331\\
CoRoT-7 & 5330\phantom{c} & 2.5 & 0.80\tablenotemark{d} & 0.225\phantom{e} & -4.802\\
Kepler-4 & 5860\tablenotemark{c} & 2.2 & 0.62\tablenotemark{d} & 0.168\phantom{e} & -4.936\\
Kepler-5 & 6300\tablenotemark{c} & 4.0 & 0.50\tablenotemark{d} & 0.148\phantom{e} & -5.037\\
Kepler-6 & 5650\tablenotemark{c} & 3.0 & 0.68\tablenotemark{d} & 0.160\phantom{e} & -5.005\\
Kepler-7 & 5930\tablenotemark{c} & 4.2 & 0.59\tablenotemark{d} & 0.155\phantom{e} & -5.099\\
Kepler-8 & 6210\tablenotemark{c} & 10.5 & 0.52\tablenotemark{d} & 0.153\phantom{e} & -5.050\\
\enddata
\tablenotetext{a}{Also known as WASP-11.}
\tablenotetext{b}{Values for \teff~(K) and \vsini~(km s$^{-1}$) were determined from a SME analysis of the Keck/HIRES spectra as described in \citet{fischer05} where available, otherwise values from the literature were used instead.}
\tablenotetext{c}{We revert to the published values for stellar effective temperature and \vsini~in these cases.}
\tablenotetext{d}{In some cases measured \bv~values were unavailable or were inconsistent with the star's spectral type.  In these cases we opted to set the \bv~color for each star equal to the value given in Table B.1 in \citet{gray05} for a given effective temperature.  We do not account for variations in \bv~due to metallicity or age, as these quantities have a negligible effect on our final \logr~values for the range of stellar spectral types present in our sample.}
\tablenotetext{e}{These hot, rapidly rotating F stars have \sval~and \logr~values that are suggestive of activity but they are near the edge of the range in \bv~for which these indices are calibrated, and their spectra show no detectible emission in the \caii~line cores.  We argue in \S\ref{rapid_rotators} that these stars are most likely chromospherically quiet.}
\tablenotetext{f}{\sval~and \logr~values are only calibrated for stars with \bv~between $0.5.-1.4$ ($4200-6200$~K).  We give values for all of the stars in our sample, but values for spectral types outside this range should be treated with some skepticism.}
\end{deluxetable}

\subsection{\sval~and \logr~for F Stars}\label{rapid_rotators}

Although we calculate \sval~and \logr~values for the F stars (\teff=$6000-6500$~K) in our sample, we note that the \logr~scaling relations described in \citet{noyes84} are not well calibrated for \bv$<0.5$.   Because F stars have higher continuum fluxes in the region of the \caii~lines, it can be difficult to detect small amounts of emission in the line cores even when it is present.  Our values for \sval~and \logr~for XO-3 (\teff=6430~K, \vsini=18.5 km s$^{-1}$) and HD 147506 (\teff=6260~K, \vsini=20.7 km s$^{-1}$) suggest that both stars are active, but these values should be regarded with some suspicion as these stars are on the edge of the calibrated range in \bv.  

As an additional test, we carried out a visual inspection of the spectra for XO-3 and HD 147506 and could find no evidence for emission in the \caii~line cores.  It is possible that the rotational broadening could be obscuring the presence of weak lines; we test this theory by searching the database of all CPS spectra and selecting a sample of F stars with effective temperatures and \sval~values similar to those of XO-3 and HD 147506 ($6000<~$\teff$~<6500$~K, $0.18<~$\sval$~<0.24$), but with significantly lower ($<7$ km s$^{-1}$) values of \vsini.  We then artificially broaden these spectra to a \vsini~of 20 km s$^{-1}$, comparable to that of XO-3 and HD 147506, using the standard IDL routine lsf\_rotate.pro, available as part of the astrolib distribution and described in \citet{gray05}.  A characteristic double-peaked emission feature is clearly detectible in the line cores of the unbroadened spectra (see Fig. \ref{vsini_spectra}), and although this signal is obscured somewhat by the broadening, there is still a noticeable flattening or even an increase in flux near the centers of the line cores in the high \vsini~versions of the spectra.  In contrast to these systems, XO-3 and HD 147506 (also plotted in Fig. \ref{vsini_spectra} for comparison) both have v-shaped spectra more characteristic of quiet stars.  We conclude that both stars are likely to be chromospherically quiet, with the additional caveat that the standard \sval~and \logr~indices used to measure this emission are not well-calibrated and are probably not appropriate to use for stars of this spectral type.

\section{Observations of Hot Jupiter Emission Spectra}\label{sec_eclipses}

There are currently four planets classified in the literature as having ``non-inverted" atmospheres, including HD 189733b \citep{deming06,grillmair07,grillmair08,charbonneau08,swain09a}, TrES-1 \citep{charbonneau05}, TrES-3 \citep{fressin10}, and WASP-4 \citep{beerer10}.  These planets have emission spectra that are consistent with standard 1D cloud-free atmosphere models where temperature decreases with increasing height in the atmosphere \citep[e.g.][]{barman08,burrows08,fortney08,madhu09}, or display at most a weak inversion \citep[e.g. WASP-4;][]{beerer10}.  Models used to describe these planets exhibit CO and H$_2$O features in absorption, which provides a good match to the shape of their observed emission spectra in the mid-IR (3.6-24~\micron).  Observations of TrES-1, TrES-3, and WASP-4  are limited to a subset of the \emph{Spitzer} IRAC bandpasses (3.6, 4.5, 5.8, 8.0~\micron), which have relatively wide ($0.75-2.9$~\micron~FWHM) transmission functions and are therefore somewhat ambiguous in their interpretation, but HD~189733b has been observed at higher wavelength resolutions with both IRS on \emph{Spitzer} \citep[$5-14$~\micron;][]{grillmair07,grillmair08} and NICMOS on the \emph{Hubble Space Telescope} \citep[$1.5-2.5$~\micron;][]{swain09a}.  Published models for HD~189733b in which the atmospheric chemistry and pressure-temperature profiles are allowed to vary in the fit consistently find that this planet is best-described by a pressure-temperature profile that decreases with height in the atmosphere \citep{madhu09,swain09a}.

\begin{figure}
\epsscale{1.2}
\plotone{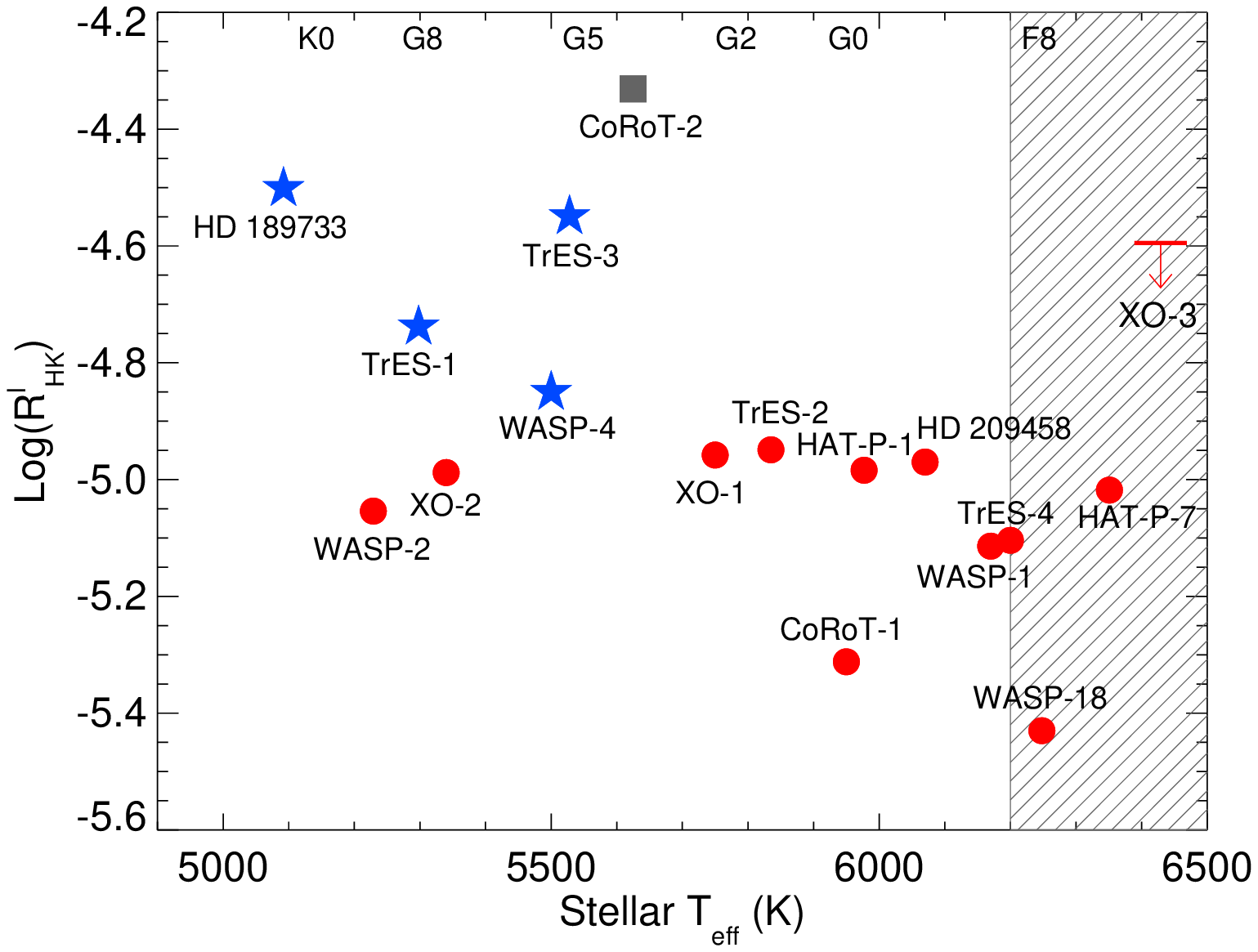}
\caption{Stellar effective temperature vs. \caii~activity index \logr, which tracks the amount of emission in the \caii~line core.  Active stars have stronger emission features and correspondingly large values of \logr; because the calibration for \logr~is uncertain for stars with effective temperatures higher than 6200~K \citep{noyes84}, we have marked this region with diagonal grey lines on this plot.  Planets classified in the literature as having temperature inversions are shown as red circles, while planets in the literature that are well-described by non-inverted atmosphere models are shown as blue stars (see \S\ref{sec_eclipses} for a discussion of WASP-2).  CoRoT-2 (grey square) has an unusual spectrum that is not well-described by either model, and XO-3 (red downward arrow) has temperature inversion and orbits a star with a relatively high \logr~value but no visible emission in the \caii~line cores by inspection (see \S\ref{rapid_rotators}).  Although cooler stars tend on average to be more active, there is still a range of activity possible within a given spectral type: XO-2, WASP-2, and TrES-1 differ by less than 100~K, and yet TrES-1 is moderately active while XO-2 and WASP-2 appear to be quiet.  For reference we give the spectral types for a range of effective temperatures at the top of the plot.}
\label{tstar_vs_activity}
\end{figure}

In contrast, the emission spectra of planets such as HD 209458b \citep{deming05,richardson07,knutson08,swain09b}, TrES-2 \citep{odonovan10}, TrES-4 \citep{knutson09}, HAT-P-1 \citep{todorov10}, HAT-P-7 \citep{christiansen10}, WASP-1b \citep{wheatley10}, WASP-18b \citep{nymeyer10}, XO-1b \citep{machalek08}, XO-2b \citep{machalek09}, XO-3b \citep{machalek08}, and CoRoT-1b \citep{deming10} are best-described by models with a stronger temperature inversion between 0.1-0.01 bar and water and CO bands in emission \citep[e.g.][]{hubeney03,fortney06,fortney08,burrows07,burrows08,madhu09,spiegel10}.  As before, our information on the majority of these systems is limited to observations in the \emph{Spitzer} IRAC bandpasses, although HD~209458b has also been observed with IRS \citep{richardson07} and NICMOS \citep{swain09b} and shorter-wavelength data is available for a few additional planets \citep{borucki09,alonso09,snellen09,christiansen10,lopez10}.  Model fits to HD 209458b's emission spectrum in which the chemistry and pressure-temperature profiles are allowed to vary find that this planet is best-described by a model with a temperature inversion at altitude \citep{madhu09,swain09b}, consistent with the conclusions of more highly-constrained models \citep[e.g.][]{burrows07}.

Although \citet{wheatley10} classify WASP-2b as a non-inverted planet, this interpretation relies on a specific model for creating the inversion (a solar metallicity atmosphere and gas phase TiO in local thermal equilibrum) that does not always provide a good match for the observed spectra of other planets \citep[e.g. HD 209458b;][]{fortney08}.  We find that this planet's emission spectrum is reasonably well-described by a simple 1840~K blackbody function, and further note that it deviates from a blackbody in ways that are more characteristic of an inverted atmosphere, with a flux that is lower than predicted in the 3.6~\micron~band and higher in the 4.5~\micron~band (see \S\ref{empirical_rel}).  For the purposes of this paper we therefore place this planet in the inverted class, and predict that more generalized models for temperature inversions \citep[e.g.][]{burrows08,madhu09} should provide an improved fit to the observed features, albeit with marginal statistical significance given the relatively large uncertainties in the planet's observed emission spectrum.

Unlike the ambiguous case of WASP-2b, CoRoT-2b has a well-measured emission spectrum that does not match the predictions of either class of models  \citep{gillon10,deming10}.  The 8.0~\micron~flux for this planet is anomalously low relative to its 4.5~\micron~flux \citep[for a discussion of possible explanations see][]{deming10}, and we therefore denote this planet in Fig. \ref{tstar_vs_activity} and \ref{empirical_hk} with a gray square rather than placing it in either of the two standard classes.  

We exclude planets with observations in only one bandpass \citep[e.g. HD~149026b;][]{harrington07,knutson09}, as well as systems with only ground-based observations \citep[e.g. WASP-19b;][]{gibson10,anderson10}, as models for these planets are poorly constrained.  We also exclude cooler ($<1000$~K), core-dominated planets such as GJ 436b and HAT-P-11b, as these planets may not fit into the simple classification scheme used to describe hot Jupiters atmospheres \citep[indeed, this appears to be the case for GJ 436b;][]{stevenson10,beaulieu10}.  We classify each of the well-characterized hot Jupiters as ``inverted" or ``non-inverted" according to the best-consensus interpretation available in the literature for that planet, except where otherwise noted.  Fig. \ref{tstar_vs_activity} shows the stellar effective temperatures \teff~and \caii~line strengths for each of these systems.  We find that the ``non-inverted" atmosphere types are consistently associated with the most chromospherically active stars, whereas ``inverted" atmosphere types are associated with quiet stars. 

\section{An Empirical Method for Classifying Hot Jupiter Emission Spectra}\label{empirical_rel}

The classification scheme described in \S\ref{sec_eclipses} relies on comparisons between the observed hot Jupiter emission spectra and 1D model atmospheres for these planets.  In this section we develop a model-independent classification scheme that relies on the relative secondary eclipse depths in the 3.6 and 4.5~\micron~IRAC bandpasses to distinguish between atmosphere types.  We select these two bands because they are (a) some of the most widely available observations for hot Jupiter atmospheres, (b) are usually the most precisely-measured eclipse depths available, and (c) provide the strongest constraints on atmosphere models \citep[e.g.][]{madhu09}.  Because the 3.6~\micron~data for TrES-1 are currently unpublished, we completed a preliminary analysis of the secondary eclipse in this channel using the methods described in \citet{knutson09} and derive an eclipse depth of $0.085\% \pm 0.013\%$, which we use for the analysis presented here.  

\begin{figure}
\epsscale{1.2}
\plotone{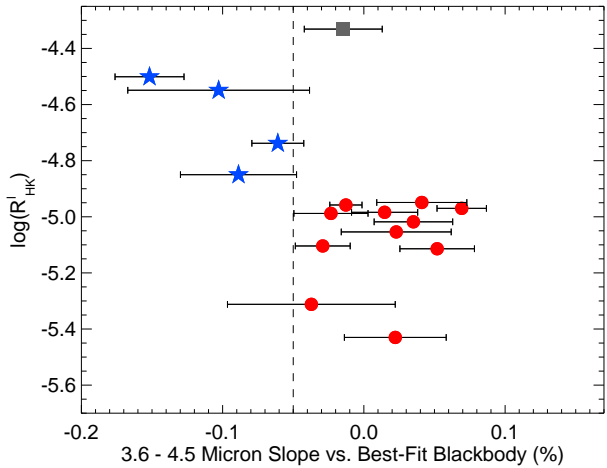}
\caption{Empirical index for classifying hot Jupiter emission spectra vs \caii~activity index \logr.  We derive our index by fitting the 3.6 and 4.5~\micron~secondary eclipse depths as measured in the \emph{Spitzer} IRAC bands with a blackbody function for the planet and take the difference between the measured slope across the 3.6 and 4.5~\micron~bands and the value predicted by the best-fit blackbody function.  Planets with blackbody emission will have an index close to zero, whereas planets with negative values are brighter than predicted at 3.6~\micron~and fainter at 4.5~\micron, and planets with positive values have the opposite behavior.  Planets classified in the literature as having atmospheric temperature inversions are shown as red circles, while planets without temperature inversions are shown as blue stars (see \S\ref{sec_eclipses} for a discussion of WASP-2b).  CoRoT-2b (grey square) has an unusual spectrum that is not well-described by either model, and we have excluded XO-3 from this plot as its activity level is ambiguously determined from its spectrum (see \S\ref{rapid_rotators}).  The vertical dashed line at -0.05 delineates a distinction between atmosphere types; in reality there is likely a continuum of behaviors ranging from the default strong absorption (i.e., no inversion), to weak absorption (corresponding to a weak inversion), to no absorption or even emission from an increasingly strong stratosphere.}
\label{empirical_hk}
\end{figure}

In order to classify each planet, we take the measured planet-star flux ratios in the 3.6 and 4.5~\micron~bands and fit them with a blackbody model for the planet where the temperature is allowed to vary freely.  We use a PHOENIX model atmosphere \citep{hauschildt99} for the star where we  interpolate in \teff~and log(g) to match the star's observed properties and set the metallicity equal to zero.  We calculate the predicted planet-star flux ratio in each of the \emph{Spitzer} bands using the transmission functions provided in the Spitzer Observer's Manual and then compare the resulting predictions to the observed slope across the $3.6-4.5$~\micron~bands.  Planets with strong inversions have lower fluxes at 3.6~\micron~and higher fluxes at 4.5~\micron~compared to our best-fit blackbody function, while planets without inversions show stronger emission at 3.6~\micron~and weaker emission at 4.5~\micron.  In the models described in \S\ref{sec_eclipses}, this difference results from the fact that the 4.5~\micron~band contains both water and CO bands while the 3.6~\micron~band is relatively unaffected by these molecules.  As these molecules switch from absorption to emission, it alters the slope of the planet-star flux ratio across these two bands accordingly.  We show the resulting difference in slopes in Fig. \ref{empirical_hk}, where ``non-inverted" atmosphere types have negative slopes and ``inverted" atmospheres have positive slopes as compared to our best-fit blackbody functions.  Planets whose emission spectra are consistent with a blackbody (i.e. no significant absorption or emission) have values close to zero in this index.

This empirical classification scheme provides a good match to the ``inverted/non-inverted" classifications presented in the literature, with the four planets with the most negative values in this index (HD 189733b, TrES-1, TrES-3, and WASP-4) corresponding to the ``non-inverted" atmosphere types described above.   Planets with significantly negative slopes across these two bands are associated with the most active stars, providing a model-independent confirmation of the proposed correlation between increased stellar activity and hot Jupiter atmosphere types.  Although \citet{wheatley10} place WASP-2b in the non-inverted class of atmospheres, we find that our index for this planet has a value of $0.025\pm0.039$, consistent with a blackbody or slightly inverted atmosphere.  As noted earlier, CoRoT-2b's anomalous emission spectrum is poorly fit by both inverted and non-inverted atmosphere models, although it is possible to fit the 3.6 and 4.5~\micron~eclipse depths with a 1790~K blackbody as shown in Fig. \ref{empirical_hk}. 

\section{The Connection Between Chromospheric Activity and UV Flux}\label{uv}

The observed correlation between hot Jupiter emission spectra and stellar activity levels would seem to suggest that the increased UV flux experienced by the planet destroys the high-altitude absorber responsible for the formation of temperature inversions.  In order to put this correlation on a more quantitative footing, we explore the connection between \caii~emission line strengths and FUV fluxes in G and K stars.  We focus our study on \lya, as this line is generally the single strongest feature in the UV spectra of late-type stars; the sun emits as much flux in \lya~as in all other wavelengths combined shortward of 1500~\,\AA~\citep{fontenla88,landsman93}.  In this exercise we compare stars of similar spectral type but differing activity levels, and we therefore elect to work with the simpler \sval~values instead of \logr~as our activity measure.  \logr~uses an empirical function of $B-V$ to remove the trend towards increasing \sval~values for later spectral types (caused by the lower continuum fluxes in the region of the \caii~lines in these stars), but this correction is unnecessary when comparing stars of the same spectral type.

It is well-established that increased emission in the \caii~line cores in late-type stars is an indicator of nonradiative heating processes in the chromosphere connected to the presence of enhanced magnetic fields \citep[e.g.,][]{leighton59,noyes84,schrijver89,bailunas95}.    Increased stellar activity as measured using \ion{Mg}{2} H \& K has also been observed to correlate with increased \lya~and x-ray emission \citep[e.g.,][]{wood05}, but to the best of our knowledge there are no quantitative scaling laws directly relating \caii~line strengths and the integrated \lya~flux as a function of spectral type for a large sample of stars.  We address this issue by selecting a sample of 26 stars with effective temperatures between $5000-6000$~K observed in \lya~by the \emph{Hubble Space Telescope} STIS instrument and its precursor GHRS as described in \citet{wood05}.  For each of these stars we obtain a measurement of the \caii~emission line strengths either from Keck HIRES observations when available, or from standard catalogues published in \citet{duncan91} and \citet{henry96}.  We also include the sun in our table, with an average \lya~flux of $5.6\times 10^{-6}$ times its bolometric flux \citep{woods00,wood05} and a \sval~value of 0.133 \citep{livingston07}.  We then bin the stars into four categories according to effective temperature ($5000-5500$~K and $5500-6000$~K bins) and activity level, where active stars are defined as those with \sval$>0.3$ for the cooler temperature bin and \sval$>0.2$ for the higher temperature bin.  We then calculate the \lya~fluxes corresponding to each bin using the values given in Table 3 in \citet{wood05}, where the authors have used the \lya~line shape information to correct for the effects of absorption from the ISM and any stellar outflows.  The UV spectra used by Wood et al. are based on a collection of archival HST observations and tend to be biased towards more active stars; our active bins have 8 cooler stars and 11 hotter stars, while the quiet star bins each contain four stars.  

We find that the average \lya~flux as a fraction of the total bolometric flux is $(1.1\pm0.2)\times 10^{-5}$ for quiet stars between $5000-5500$~K and $(4.2\pm0.2)\times 10^{-5}$ for active stars in that same temperature range, where the quiet stars have an average \sval~of 0.171 and the active stars have an average value of 0.545 and we set the uncertainty on the \lya~flux estimates equal to the standard deviation of the \lya~fluxes for the stars in that bin.  This range in \sval~is comparable to the observed difference between HD 189733 (5100~K, \sval$=0.508$) and its closest quiet analogue WASP-2 (5230~K, \sval$=0.159$).  By analogy, then, we expect that HD 189733 will show an enhancement in \lya~flux of approximately a factor of four relative to WASP-2.  Because the two planets in these systems orbit at nearly identical distances from their host stars (0.031 A.U. in both cases), we expect the incident UV flux at the surface of HD~189733b to be a factor of four higher than at the surface of WASP-2b.  

We repeat this same experiment for the $5500-6000$~K temperature bins and find an average \lya~flux of $(5.0\pm0.6)\times 10^{-6}$ times the bolometric flux for quiet stars and $(3.6\pm1.1)\times 10^{-5}$ times the bolometric flux for active stars, corresponding to a factor of eight enhancement for an increase in average \sval~from 0.157 to 0.354.  TrES-3 has an effective temperature of 5530~K and \sval=0.305, while its closest quiet analogue XO-1 has an effective temperature of 5750~K and \sval=0.168.  Taking into account the planet TrES-3b's slightly smaller orbital distance (0.023 A.U. vs. XO-1b's 0.049 A.U.), we estimate that it experiences a \lya~flux approximately thirty times higher than that experienced by XO-1b.  HD 209458 is the only planet-hosting star with a direct measurement of its \lya~flux in \citet{wood05}, and has a total flux of $6.9\times 10^{-6}$ times its bolometric flux, consistent with the other quiet stars in the $5500-6000$~K bin.

WASP-4 (5500~K) falls on the boundary between temperature bins, but it is indistinguishable from TrES-3 (5530~K) and so we estimate its \lya~flux using the hotter temperature bin.  If we assume the \lya~flux scales linearly with \sval, WASP-4 (\sval$=0.194$) would have a \lya~flux of $1.1\times 10^{-5}$ times its bolometric flux.  Because the planet WASP-4b orbits at a comparable distance to that of TrES-3b, this means that WASP-4b would receive approximately one third the flux of this planet, ten times higher than the flux experienced by XO-1b.  Comparing it to the stars in the cooler temperature bin, WASP-4b receives approximately half as much \lya~flux as HD~189733b and twice as much as WASP-2b.

According to the empirical index plotted in Fig, \ref{empirical_hk}, TrES-1b lies closest to the boundary between inverted and non-inverted atmospheres, and this is reflected in its UV flux.  The star TrES-1 (5300~K) has a \sval~of 0.244, corresponding to a \lya~flux of $1.1\times 10^{-5}$ times its bolometric flux.  This flux is comparable to that of WASP-4, but the planet TrES-1b orbits at a greater distance and as a result the incident flux at the surface of the planet is much less, only $30\%$ greater than that received by WASP-2b.  

Hot Jupiter atmospheres have a relatively high absorption cross section in \lya~and we expect that much of this incident flux will be absorbed in the upper regions of the atmosphere, where relative abundances of molecules may differ from local thermal equilibrium by many orders of magnitude.  Because photolysis cross sections and return reaction rates are species-specific and can encompass multiple pathways, it is difficult to predict a priori the effects that a significantly increased \lya~flux would have on the relative abundance of the unknown stratospheric absorber.  We suggest that this would be best addressed by comprehensive photochemistry models such as those described in \citet{zahnle09,zahnle10} and \citet{line10}.  

\section{Discussion \& Conclusions}\label{discussion}

We measured the \caii~line strengths for fifty transiting planet host stars and correlated these activity indicators with planet atmosphere type.  Planets with non-inverted atmospheres are found orbiting the most chromospherically active stars, while planets with inverted atmospheres are found orbiting quiet stars.  The observational correlation described above is independent of the inverted/non-inverted classification scheme for hot Jupiter atmospheres, as we find the same result for an empirical classification scheme based on 3.6 and 4.5~\micron~secondary eclipse depths alone.  We estimate the statistical significance of our result using a rank order test in which we randomly select four stars out of our sample of fifteen systems observed with both Keck/HIRES and \emph{Spitzer} after excluding XO-3 and CoRoT-2b and ask whether those four stars have higher \logr values than the remaining eleven stars.  We then repeat this test for a million different trials and find that we obtain the four largest \logr values for our non-inverted planets by chance only 0.13\%~of the time, corresponding to a $3.4\sigma$ result where the significance is limited by the small size of our sample.

We find that active G and K stars likely exhibit an enhancement of $4-7$ in their \lya~emission relative to quiet stars with similar effective temperatures, and suggest that the increased UV flux from active stars destroys the high-altitude absorber responsible for the formation of temperature inversions.  This theory is consistent with the sulfur model proposed by \citet{zahnle09}, but could also be applied to a range of other scenarios.  Because \caii~and UV emission are only indirectly linked, definitive confirmation of this theory will require an estimate of the UV fluxes incident at the surfaces of these planets, taking into account both the spectral types and activity levels of their host stars as well as their orbital distances.  Orbital eccentricity may also prove to be important, as planets such as XO-3 likely exist in a pseudo-synchronous spin state in which incident radiation is distributed more evenly over the surface than in the tidally locked case.  If this is true, eccentric planets may require higher UV fluxes in order to suppress their dayside temperature inversions as compared to their tidally locked counterparts.  We will address this issue in more detail in a follow-up paper, in which we examine the relationship between spectral type, \caii~emission, and \lya~emission and derive individual UV flux estimates for the transiting planets listed in Table \ref{ca_hk_table}.    

Although \emph{Spitzer} exhausted the last of its cryogen in May 2009, the 3.6 and 4.5~\micron~channels on IRAC are still functioning at full sensitivity and the telescope continues to operate as part of an extended warm mission using those two channels.  As part of this warm mission we are currently executing a program to observe secondary eclipses of 19 additional transiting planets not observed during the cryogenic mission (ES program 60021, PI Knutson).  When combined with the remaining unpublished secondary eclipse data from Spitzer's cryogenic mission and other ongoing warm mission programs (e.g. ES 60028, PI Charbonneau; DDT 60003, PI Harrington), data for $20-30$ additional transiting planet systems  should become available by  summer 2011.  We propose that the empirical scheme described in \S\ref{empirical_rel} can be used to provide an initial classification for these new systems, while the increasing availability of ground-based secondary eclipse detections as described above should help to provide better constraints for atmosphere models in cases where we are limited to the two shorter-wavelength \emph{Spitzer} bands.  Because active stars are less common in the sample of transiting planets (perhaps due to selection biases among the ground-based photometric surveys and the increased difficulty of radial velocity follow-up work for such systems), we suggest that the analysis of secondary eclipse data for planets orbiting active stars be given a high priority within the current large ongoing \emph{Spitzer} programs. 

\acknowledgments

We would like to thank Geoff Marcy and Debra Fischer for their generous assistance in obtaining the Keck/HIRES spectra and deriving the SME values used in this analysis, as well as Lucianne Walkowicz, Kevin Zahnle, Mark Marley, Jonathan Fortney, Drake Deming, Jean-Michel Desert, and Adam Burrows for providing invaluable commentary on various aspects of this project.  HAK is supported by a fellowship from the Miller Institute for Basic Research in Science.  We thank the many observers who contributed to the data reported here.  We gratefully acknowledge the efforts and dedication of the Keck Observatory staff, especially Scott Dahm, Hien Tran, and Grant Hill for support of HIRES and Greg Wirth for support of remote observing.  A.\,W.\,H.\ gratefully acknowledges support from a Townes Post-doctoral Fellowship at the U.\,C.\ Berkeley Space Sciences Laboratory.  Finally, the authors wish to extend special thanks to those of Hawai`ian ancestry on whose sacred mountain of Mauna Kea we are privileged to be guests.  Without their generous hospitality, the Keck observations presented herein would not have been possible.

\end{document}